# A 4-Element MIMO Baseband Receiver with >35dB 80MHz Spatial Interference Cancellation

Erfan Ghaderi[1], *Student Member*, *IEEE*, Ajith Ramani[2], *Member*, *IEEE*, Arya Rahimi[1], *Student Member, IEEE*, Sudip Shekhar[2], *Senior Member, IEEE* and Subhanshu Gupta[1], *Senior Member*, *IEEE*

*Abstract*— Next-generation communication systems with wide bandwidths need to operate in interference-limited networks. A discrete-time delay (TD) technique in a baseband receiver array is proposed for canceling wide modulated bandwidth spatial interference and reducing the ADC dynamic range requirements. The proposed discrete TD technique first aligns the interference using non-uniform sampled phases followed by uniform cancellation using a Truncated Hadamard Transform implemented with antipodal binary coefficients. A digital time-interleaver with 5 ps resolution spanning 15 ns implements a scalable discrete TD to compensate the inter-element delay, while the multiply-accumulate in the signal path is simplified by implementing a 1-bit differential truncated Hadamard matrix. Measured results demonstrate greater than 35 dB cancellation over 80 MHz modulated bandwidth in 65 nm CMOS with a 592× improvement over prior-art demonstration of wide modulated bandwidth interference cancellation.

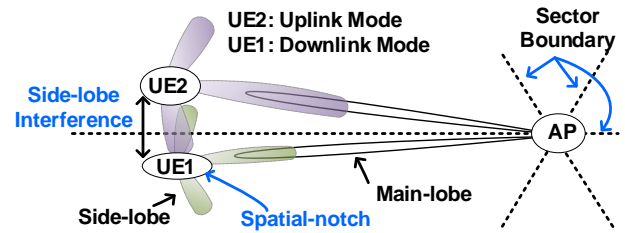

Fig. 1. Illustration of interference in dense small cells with aggressive frequency use. User elements UE1 and UE2 use the same frequency channel across the sector boundaries.

## I. INTRODUCTION

The exponential growth in wireless communication is spurring development of Multiple-Input and Multiple-Output (MIMO) transceivers for higher spectral efficiencies, wide modulation bandwidths (BWs), and spatial multiplexing gains. However, as next-generation sub-6GHz and millimeter-wave networks [1] progress towards ultra-dense small cells, the dramatic shortening of inter-cell distances will result in both line-of-sight and non-line-of-sight interference channels [2]-[4]. This problem is severe for multi-user communications networks operating in crowded electromagnetic environments and can significantly degrade the performance of radios.

An example scenario in Fig. 1 shows two user elements (UEs) located in adjacent cell-sectors communicating with an access point (AP) in uplink and downlink modes respectively. Even though the strength of side-lobe is only 10-15 dB lower than the main-lobe, the in-band (same frequency channel) side-lobe interference can desensitize the receiver (RX) or degrade the in-band SNR if not addressed. This interference problem worsens near the sector boundary creating poor connection zones, especially if the small cell is operating near maximum capacity. Multiple system parameters impact the onset of an interference-limited network behavior over noise-limited behavior [3]-[5]. These parameters can include the number of UEs and APs, antenna gains, operational BW, blockage characteristics, and the choice of MIMO architecture. Though frequency allocation schemes have been proposed to mitigate interference [6], [7], the next-generation communication networks require considerable frequency planning and integrated spatial cancellation techniques.

While interference mitigation using electronic phased-arrays has been an active research area [8-14], the dense small-cell mobile networks with wide modulated BWs and higher data rates for next-generation systems impose unique challenges to the front-end hardware that has not been addressed in prior art. The phase-shift (PS) method employed for interference cancellation in prior-art, being an approximation of a time-delay (TD) cancels the interference at a single frequency (i.e., generates a single null). For next-generation systems with wide modulated BWs, it results in interference leakage and significantly higher dynamic range requirements for the baseband (BB) and the ADC [2]. The PS based approximation (and hence, leakage) can be corrected by augmenting the RX front-end with a true TD (TTD) in BB as described in Section II. However, such an approach needs precise delay generation and compensation. This work presents the first interference cancellation technique for spatially separated wide modulated BW signals that coexist in the same frequency channel using discrete TD augmented RX front-ends.

In our prior work [15], we introduced a high range-to-resolution ratio BB TTD for frequency-uniform beamforming gain to overcome the beam squinting problem in PS-based beamformers. In this paper, we leverage the BB delay-compensating technique [15] with Truncated Hadamard Transform (THM) to implement a 4-element wide modulated BW Spatial Interference Cancellation (SpICa) BB RX array. Our contributions in this paper compared to our previous work [15] are as follows:

a. SpICa for dense interference-limited networks, where only beamforming cannot support the multiple wide modulated BW channel coexistence, is developed.

Manuscript submitted July 1st, 2019. This research work was partly supported by National Science Foundation (NSF) award #1705026 and Natural Sciences and Engineering Research Council of Canada (NSERC).
[1]E. Ghaderi, A. Rahimi and S. Gupta are with the School of Electrical Engineering and Computer Science, Washington State University, Pullman, WA, USA. (erfan.ghaderi@wsu.edu).
[2]A. Ramani and S. Shekhar are with the Department of Electrical and Computer Engineering, University of British Columbia, Vancouver, BC, V6T1Z4, Canada.

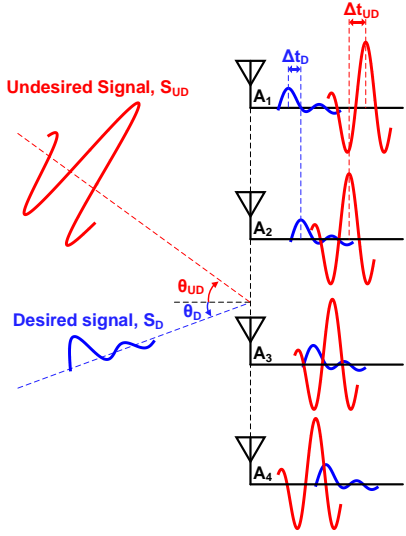

Fig. 2. Time domain representation of the source and received desired and undesired signals in a four-element linear antenna array.

b. A wide modulated BW SpICa technique in BB is implemented through delay-compensating and THM for the first time.
c. Detailed circuit implementations of the delay-compensating technique and its application for wide modulated BW SpICa in a 4-element BB RX array are presented.
d. SpICa measurements that demonstrate significant performance improvement over prior-art are presented.

A brief overview of SpICa techniques is presented in Section II. An analytical study of interference leakage in PS-based schemes is conducted to understand the impact of multiple system parameters in an interference-limited network behavior necessitating the need for TTD-based SpICa. Section III describes the design of the proposed time-interleaved staggered switched-capacitor array implementing discrete TD based SpICa. Section IV details the measured results for a four-element BB RX array. Section V concludes the proposed work.

## II. STATE-OF-ART SPICA AND NEED FOR TD WIDE MODULATED BW SPICA

This section studies state-of-art SpICa schemes in the presence of multiple strong and weak single-tone, narrowband (NB) or wide modulated BW spatially-distinct in-band signals [8]-[11]. These schemes are needed to avoid the need for large BB and ADC dynamic range. We will further motivate the need of a TTD-based SpICa for wideband modulated signals.

The non-zero spacing of the antennas leads to a delay between the received signals at each antenna. We model this phenomena in a phased-array system as $y(t)=\Gamma_D[S_D(t)] + \Gamma_{UD}[S_{UD}(t)]$, where $y(t)$ is the received signal vector at the antennas, $\Gamma_D$ and $\Gamma_{UD}$ are the TD generating function for the desired signal $S_D(t)$ and the undesired signal $S_{UD}(t)$, respectively. A four-element linear antenna array is shown in Fig. 2. In this array the incident input signal can be written in the time-domain as:

$$y_i(t)=S_D[t-(i-1)\Delta t_D]+S_{UD}[t-(i-1)\Delta t_{UD}], i=1...4 \quad (1)$$

where $i$ is the element number. The same equation can be described in the frequency domain as:

$$y_i(j2\pi f)=S_D(j2\pi f)\cdot e^{-j(i-1)2\pi f\Delta t_D} + S_{UD}(j2\pi f)\cdot e^{-j(i-1)2\pi f\Delta t_{UD}} \quad (2)$$

where $\Delta t_D$ and $\Delta t_{UD}$ are the delays for the desired and undesired signal, respectively, and defined as:

$$\Delta t_D = \frac{d}{\lambda_C}\cdot\frac{\sin\theta_D}{f_C} \quad (3)$$

$$\Delta t_{UD} = \frac{d}{\lambda_C}\cdot\frac{\sin\theta_{UD}}{f_C} \quad (4)$$

Here, $d$, $\lambda_C$, and $f_C$ are the element spacing, carrier wavelength and frequency respectively.

### A. PS-based SpICa schemes and interference leakage

In a PS-based MIMO RX the input signal BW is assumed to be small and (2) can be rewritten as (5):

$$y_i(j2\pi f)=S_D(j2\pi f)\cdot e^{-j(i-1)\Delta\varphi_D} + S_{UD}(j2\pi f)\cdot e^{-j(i-1)\Delta\varphi_{UD}} \quad (5)$$

where the phase shift elements $\Delta\varphi_D$ and $\Delta\varphi_{UD}$ are calculated as follows:

$$\Delta\varphi_D=2\pi f_C\cdot\Delta t_D=\frac{d}{\lambda_C/2}\pi\cdot\sin\theta_D \quad (6)$$

$$\Delta\varphi_{UD}=2\pi f_C\cdot\Delta t_{UD}=\frac{d}{\lambda_C/2}\pi\cdot\sin\theta_{UD} \quad (7)$$

This approximation is valid as long as the received signal is a single-tone or NB signal. In PS-based SpICa schemes [8]-[11], the received signals are phase shifted and properly added to cancel the undesired signal (Fig. 3). Mathematically, each $y_i(j2\pi f)$ is multiplied by $e^{j(i-1)\Delta\varphi_{UD}}$, resulting in the overall phase for the undesired signal as $\Delta\varphi_{UD} - 2\pi f\Delta t_{UD}$. This term is frequency dependent and zero at a single frequency only, leaving a residue for other frequencies offset from carrier frequency, $f_C$. This residue leads to interference leakage leading to increased dynamic range for the BB and ADC after down-conversion.

In [8], a SpICa RX is demonstrated using BB-to-RF impedance translation that models an inverse function of the incoming array pattern. Although capable of single-tone SpICa when swept over 320 MHz, SpICa for modulated interferer is not demonstrated. Other techniques include switched-capacitor vector-modulation based phase shifters [10], [11], but SpICa is achieved only for NB interferers or modulated interferer (16-QAM) over ~135 kHz [11]. We study interference leakage in a PS-based SpICa with 4, 16 and 64 elements. In these cases, we assume the even elements of $y_i(t)$ are subtracted from its odd elements. Fig. 4 shows the rejection performance of the undesired signal across the normalized frequency with angle of arrival (AoA)=45°. As it can be seen, for higher number of antennas the undesired signal leakage is larger (higher conversion gain for $S_{UD}$) and cancellation starts to become imperfect even in narrower BWs as the number of antennas are progressively increased. This simulation shows the need for a TTD implementation to compensate the delay between the received signals before performing SpICa, ideally independent of the input signal BW.

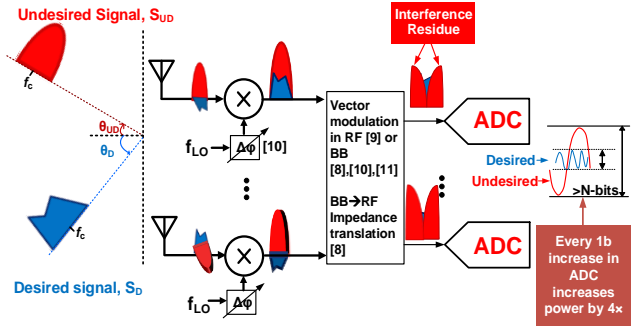

Fig. 3. PS-based SpICa (implemented using vector modulation technique) causes leakage and requires higher dynamic range ADC for wide modulated BW interferers.

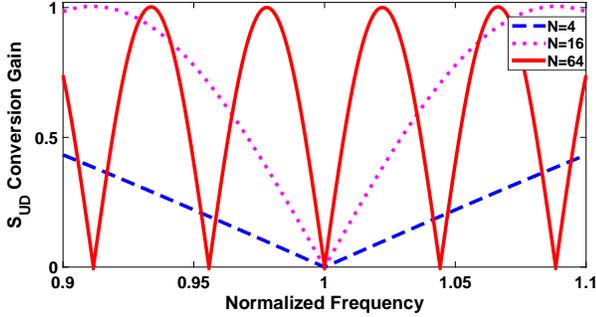

Fig. 4. SpICa simulation in a PS-based scheme vs. normalized frequency over 20% fractional bandwidth with different number of antennas.

### B. TTD based interference cancellation for wide modulated BW signals

The TTD element can be implemented at RF. In this case the delay elements, such as LC or transmission lines, can compensate the TD between the undesired signals and align them. However, the area/power hungry implementation and passive device mismatches at RF makes this solution less attractive. TTD was recently implemented digitally in [14], but the post-ADC implementation requires large ADC dynamic range in absence of any spatial interference cancellation techniques. The digital implementation for a 16-element TTD with 16 oversampled ADCs also consumes 453 mW (16 mW per ADC, 12.3 mW per element for digital TTD beamforming) for 100 MHz BB bandwidth. In a downconversion RX, RF TD implementation is mathematically identical to TD introduced in LO together with BB [15]. The TD element in LO, operating at a single frequency is simply analogous to a phase shift. Consequently, augmenting this PS in LO with a BB TD element can compensate the delay between the received signals, which is the proposed implementation in this work.

In other words, a BB delay implementation can remove the frequency-dependent residue phase ($\Delta\varphi_{UD} - 2\pi f \Delta t_{UD}$) that exists in a PS-based SpICa scheme. After time-aligning the undesired signals at the BB, SpICa can be performed through a differential orthogonal matrix, which subtracts half of the signals from the other half uniquely. At the output of the matrix, the undesired signal component is uniformly rejected across BW without any residue compared to the PS-based SpICa, i.e. the proposed intervention is angle- and frequency-independent.

In Fig. 5, the proposed BB discrete TD implementation for a four-element SpICa RX is shown. In this four-element array, there are three unique combinations of the time-aligned BB signals to perform SpICa. These three unique combinations construct a 4×3 matrix with only +1 and -1 entries, which forms a THM [16]. The received signals, including both desired and undesired, at the antenna are downconverted, phase shifted, and then fed to BB to perform TTD and SpICa. Because the SpICa is being done at BB, the RF-FE must have sufficient linearity to downconvert the strong undesired signal without distorting the desired signal, similar to requirements in prior-art BB SpICa [8], [12]. Note that the BB linearity is relaxed due to the interference cancellation prior to addition. Further RF-FE impairments, especially amplitude and phase mismatches between the elements must be taken into design considerations so as to not limit the performance of BB SpICa techniques. These mismatches can be reduced using digital techniques similar to improving the image rejection ratio in a quadrature receiver for intra-band carrier aggregation [17], [18].

### C. Effect of SpICa on wide modulated BW desired signal in a TTD-based implementation

Since the TTD implementation at LO in conjunction with BB is mathematically identical to RF TTD implementation, we examine the effect of the SpICa on the desired signal in RF domain. At the output of the THM ($OUT_i(t)$, $i$=1, 2, 3), the undesired signal is eliminated, and the residue signal (which is only a function of the desired signal) can be written as:

$$\mathbf{OUT}(t) = \begin{bmatrix} OUT_1(t) \\ OUT_2(t) \\ OUT_3(t) \end{bmatrix} = \begin{bmatrix} +1 & -1 & +1 & -1 \\ +1 & +1 & -1 & -1 \\ +1 & -1 & -1 & +1 \end{bmatrix} \begin{bmatrix} S_D[t] \\ S_D[t+(\Delta t_{UD} - \Delta t_D)] \\ S_D[t+2(\Delta t_{UD} - \Delta t_D)] \\ S_D[t+3(\Delta t_{UD} - \Delta t_D)] \end{bmatrix} \quad (8)$$

This equation can be rewritten in the frequency domain, as in (9), and desired signal conversion gain vector, $\mathbf{G_D}(j2\pi f)$, can be defined as (10):

$$\mathbf{OUT}(j2\pi f) = \begin{bmatrix} OUT_1(j2\pi f) \\ OUT_2(j2\pi f) \\ OUT_3(j2\pi f) \end{bmatrix} = \begin{bmatrix} +1 & -1 & +1 & -1 \\ +1 & +1 & -1 & -1 \\ +1 & -1 & -1 & +1 \end{bmatrix} \begin{bmatrix} S_D(j2\pi f) \\ S_D(j2\pi f)e^{j2\pi f(\Delta t_{UD}-\Delta t_D)} \\ S_D(j2\pi f)e^{j4\pi f(\Delta t_{UD}-\Delta t_D)} \\ S_D(j2\pi f)e^{j6\pi f(\Delta t_{UD}-\Delta t_D)} \end{bmatrix} \quad (9)$$

$$\mathbf{G_D}(j2\pi f) = \frac{\mathbf{OUT}(j2\pi f)}{S_D(j2\pi f)} = \begin{bmatrix} G_{D1}(j2\pi f) \\ G_{D2}(j2\pi f) \\ G_{D3}(j2\pi f) \end{bmatrix} =$$

$$\begin{bmatrix} 1 - e^{j2\pi f(\Delta t_{UD}-\Delta t_D)} + e^{j4\pi f(\Delta t_{UD}-\Delta t_D)} - e^{j6\pi f(\Delta t_{UD}-\Delta t_D)} \\ 1 + e^{j2\pi f(\Delta t_{UD}-\Delta t_D)} - e^{j4\pi f(\Delta t_{UD}-\Delta t_D)} - e^{j6\pi f(\Delta t_{UD}-\Delta t_D)} \\ 1 - e^{j2\pi f(\Delta t_{UD}-\Delta t_D)} - e^{j4\pi f(\Delta t_{UD}-\Delta t_D)} + e^{j6\pi f(\Delta t_{UD}-\Delta t_D)} \end{bmatrix} \quad (10)$$

As seen in (10), the desired signal is affected by a known frequency-dependent profile that can be equalized after digitization (Fig. 5). The transfer function profile for each of the signal paths in (10) depends on the difference between desired and undesired time delay, which itself is a function of the AoA of both desired and undesired signals. Fig. 6 plots these conversion gains vs. normalized frequency, for a signal occupying a BW equivalent to 20% of the center frequency. In this simulation, for the AoA$_D$=0° and AoA$_{UD}$= 45°, it can be seen that the conversion gains for almost the entire band are more than one and their frequency dependency profiles can be compensated by a digital equalizer post ADC.

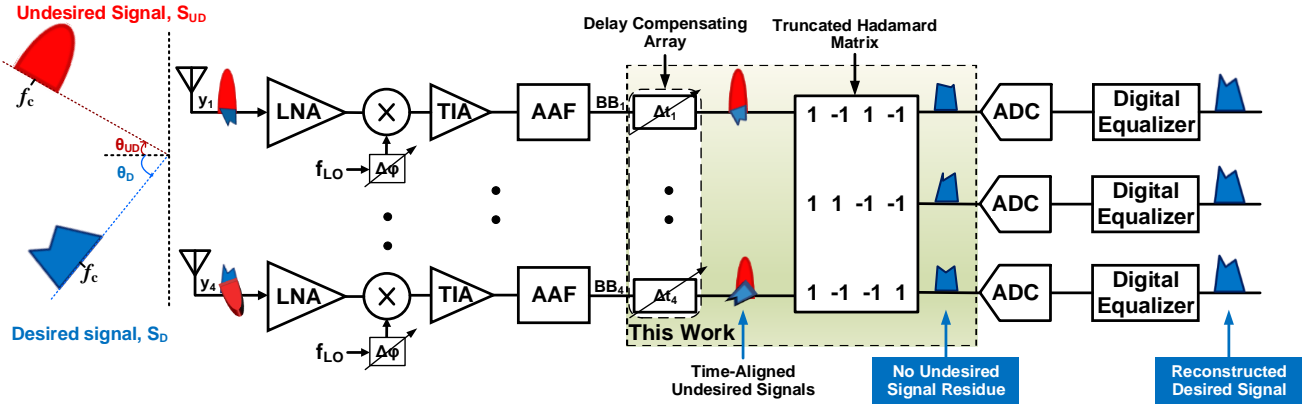

Fig. 5. Proposed discrete TD scheme with THM for wide modulated BW SpICa.

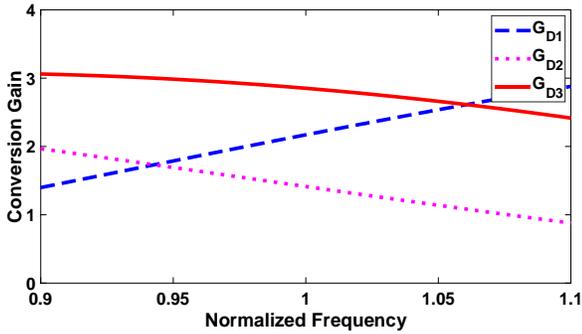

Fig. 6. Desired signal conversion gain vs. normalized frequency over 20% fractional BW in the proposed SpICa scheme.

## III. DISCRETE TD-BASED SPICA CIRCUIT DESIGN

The proposed BB wide modulated BW SpICa solution comprises of two main blocks: (i) the delay compensation array; and (ii) the cancellation matrix, which in this case is THM. The delay compensation array is implemented through a non-uniform discrete TD technique. The THM entries are simply constructed through differential implementation and the row's addition is done through charge domain summation. The proposed solution, as shown in Fig. 5, implements a BB discrete TD-based SpICa over 100 MHz modulated BW in BB. A delay compensation range of 15 ns (between the first and last antenna) and a delay compensation resolution of 5 ps is chosen for our proof-of-concept 4-element prototype. As explained later in this section, this design is highly scalable in terms of array size, SpICa performance, and BW.

### A. Discrete TD Technique

Fig. 7 shows the non-uniform sampling technique used to implement the discrete TD array and compensate the undesired TD between BB signals. Intuitively, sampling the BB signals with time delayed clocks, is equivalent to sampling the time delayed version of those signals with same clock. Using this discrete TD technique, the BB delay compensation array is realized. The resolution of the delay-compensation elements, which in this discrete TD structure is determined by the resolution of the delay between the sampling clocks, determines the SpICa performance. The longest delay sets the required level of interleaving. As an example, if 8 elements are implemented, the delay between the first and the last antenna increases to 35 ns (assuming all the other parameters are not changing). To compensate this new total delay, instead of changing the sampling rate that is determined by the signal bandwidth, we scale the level of interleaving to 8. By doing so, there will be 8 sets of sampling phases and each set consists of 8 interleaved phases.

Fig. 8(a) shows the proposed time-interleaver topology to obtain a precision of 5 ps and a maximum range of 15 ns, between the first and the last antennas, in the sampling clock phases. The time-interleaver resolution determines the theoretical maximum interference rejection. In the proposed design, Nyquist sampling results in a sampling clock frequency 200 MHz (twice of BB BW, 2×100 MHz). The external 800 MHz single-phase clock ($f_{clk}$) is first terminated with a 50 Ω on-chip resistor and is fed to a quadrature phase generator circuit [19] after amplification to rail-to-rail swing. The quadrature phase generator provides quadrature outputs ($I^-$, $I^+$, $Q^-$, $Q^+$) at 200 MHz with complete cycle coverage (=360°) and period of $T_S$ = 5 ns. These quadrature phases are fed to an 8-bit-binary is capable of 5 ps resolution (=1.25 ns/256) and has a maximum 5 ns range after the four differential quadrature outputs of the PI ($I_o^-$, $I_o^+$, $Q_o^-$, $Q_o^+$) are selected by digital phase interpolator (PI) which is programmed using an on-chip serial-to-parallel (SPI).

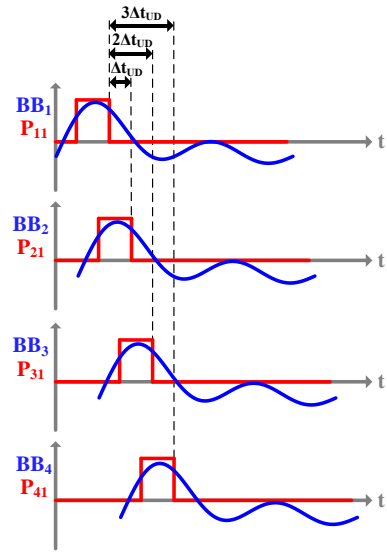

Fig. 7. Basic idea behind non-uniform sampling used to implement TD.

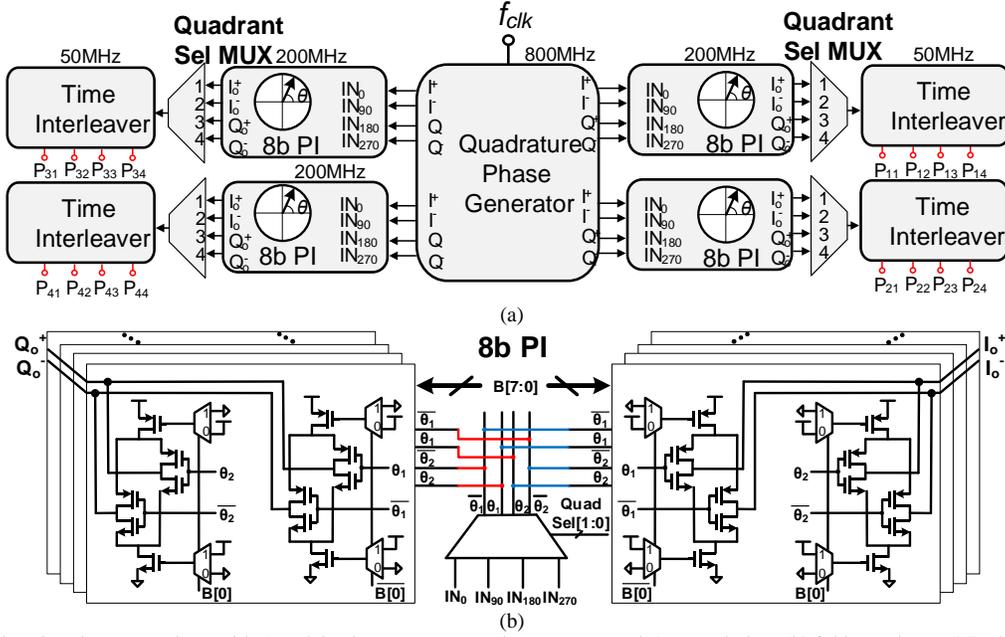

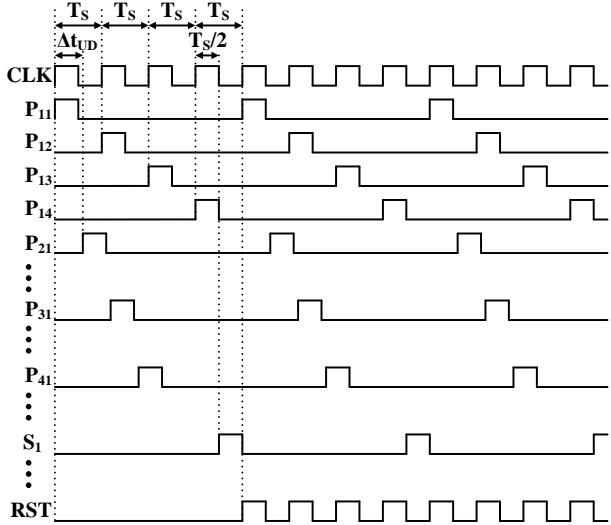

Fig. 8. (a) Time-interleaver topology with 5 ns delay between consecutive antennas and 5 ps resolution; (b) 8-bit quadrature PI schematic.

Fig. 9. Required phases to span 15 ns range and ensure proper functionality of the discrete TD-based SpICa implementation.

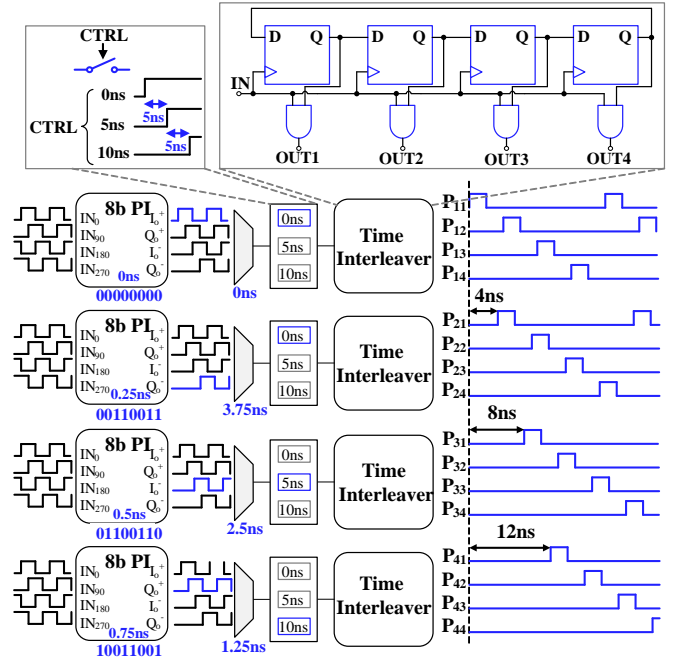

Fig. 10. States of the time-interleaver and important phases for delay compensation of 4 ns between consecutive antennas.

The period of the 200 MHz clock is insufficient to cover the required time span of 15 ns between the first and the last antenna. This is remedied by generating four phases ($P_{11}$, …, $P_{14}$), from each Q-MUX output, for each quadrant select MUX, at 50 MHz with a 12.5% ON-time from a time interleaver. These 16 phases ($P_{11}$, …, $P_{44}$) provide the required staggered-time-interleaved clocks as shown in Fig. 9. A digital implementation with power consumption proportional to $C \cdot VDD^2 \cdot f_{clk}$ permits scaling with CMOS process, VDD and BW. A finer resolution is preferred in improving the consecutive antenna, resulting in 15 ns of overall range. Such a large range is again chosen to prove that this design can be used in larger arrays, where the delay between received signals at the first and last antenna can be a large value.

In Fig. 10, the state of each PI and MUX is illustrated to generate a delay compensation of 4 ns between consecutive antennas. Each of the four PIs is configured independently using the SPI data bits. The first PI does not interpolate any of the phases generated by the quadrature phase generator preceding the PI. As such, the $I_o^+$ phase is selected by the MUX and fed to the time-interleaver to generate the first set of sampling phases ($P_{11}$, .., $P_{14}$). These sampling phases are then applied to the first input signal ($BB_1$ in Fig. 11). In order to generate 4 ns of delay for the 2$^{nd}$ antenna element, the 2$^{nd}$ PI interpolates the input clocks by 0.25 ns. Selecting the $Q_o^-$ phase of the PI by MUX results in 4 ns relative delay (0.25 ns from the PI and 3.75 ns from choosing the $Q_o^-$ phase). In order to generate 8 ns of delay for the 3$^{rd}$ antenna element, the 3$^{rd}$ PI interpolates the input clocks by 0.5 ns. Selecting the $I_o^-$ phase of the PI by MUX results in 3 ns relative delay (0.5 ns from the PI and 2.5 ns from choosing the $I_o^-$ phase). Another 5 ns (= $T_S$) of desired relative delay is introduced by enabling the time interleaver only after 5 ns relative to the first two time-interleavers. This relative delay can only be controlled as 0, $T_S$, or $2T_S$, in our implementation. Finally, a delay of 12 ns is implemented for the 4$^{th}$ antenna element.

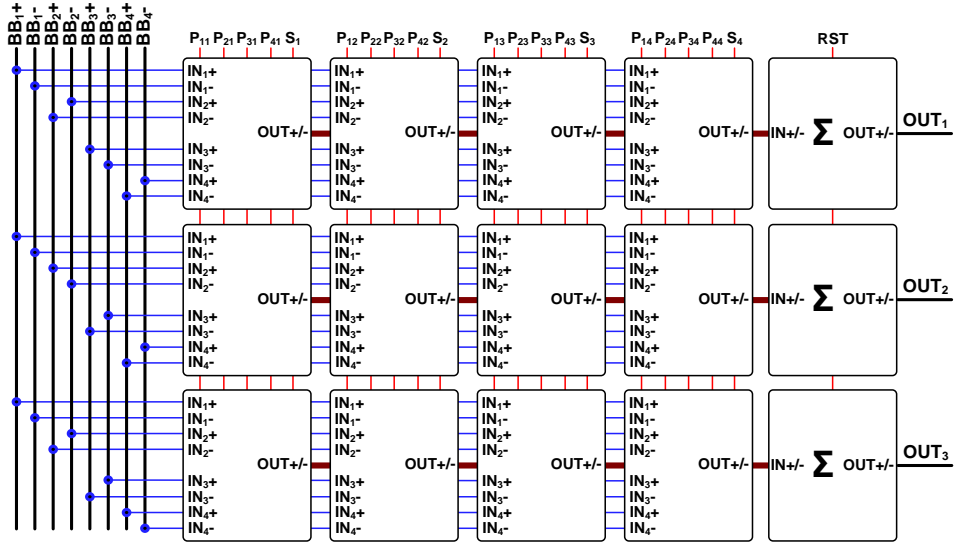

Fig. 11. THM realization through differential analog MAC implementation.

### B. THM implementation through time-interleaved switched capacitor sampling

In Fig. 11 an analog implementation of the THM is proposed. In this structure, each row of the THM is implemented as an analog multiply-accumulate (MAC), with four multipliers ($M_1 \cdots M_4$) and one accumulator. Differential implementation is used to simplify THM entries' realization. By simply flipping the polarity of the corresponding inputs to $-1$, the THM can be realized without any extra hardware requirements. Multiplications are achieved using the conventional bottom-plate sampling switched-capacitor circuit and accumulation is achieved using a fully-differential operational transconductance amplifier (OTA) based parasitic-insensitive summer (Fig. 12). Each accumulator sums the inputs, delayed and sampled on the four capacitors during the SUM phases ($S_1 \cdots S_4$). This switched-capacitor based implementation of a BB RX requires four phases for sampling ($P_{1i} \cdots P_{4i}$) followed by summation ($S_i$) and reset (RST) phase, in a four-element array (also see Fig. 9). In the sampling phases, ($P_{1i} \cdots P_{4i}$), input signal from each RX is first sampled on a sampling capacitor ($C_S$) uniquely. The input sampler is implemented using a PMOS switch optimized to provide the maximum linearity to handle input signals between 0.4 V - 1 V. The value of $C_S$ is determined by the noise requirements of the RX. After the last sampling phase ($P_{4i}$), the stored charges on each capacitor are shared in $S_i$ phase. This charge sharing performs an averaging function. To change this functionality to summation, the shared charges are transferred to the feedback capacitor ($C_F$) in a switched-capacitor summer. The OTA used in this adder must satisfy SNR and BW requirements of the RX. In the RST phase, there is a feedback network on the OTA, with feedback coefficient of 1/4 (consisting of 4 effective sampling capacitors and one feedback capacitor). Considering these requirements, the OTA is implemented as a two-stage internally compensated structure with Miller compensation (Fig. 13). The OTA is designed with more than 70 dB open loop gain, 685 MHz unity gain bandwidth, and 72° phase margin. These values ensure that the desired cancellation performance is met across the PVT variations. Because the input nodes of the OTA do not vary significantly (virtual ground) and the output nodes just carry the weaker desired signals (the undesired has been rejected), telescopic cascode structure has been used for both stages. A wide-swing cascode current mirror is used that mirrors the input off-chip bias current (=200 µA) to each of the three OTAs consuming 2 mA at 1V supply. The BW of the common-mode feedback loop is set to be greater than Nyquist frequency to allow first-order rejection of common-mode noise and interference. Note that the interference cancellation happens prior to summation resulting in significantly reduced linearity requirements for the summer.

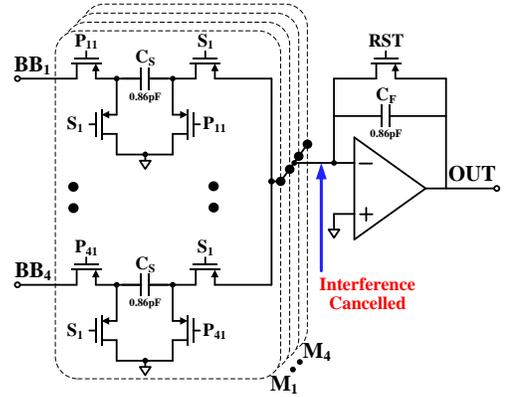

Fig. 12. Simplified single-ended representation of THM rows' implementation. Note that the interference cancellation is done at the virtual ground resulting in significantly relaxed linearity requirements.

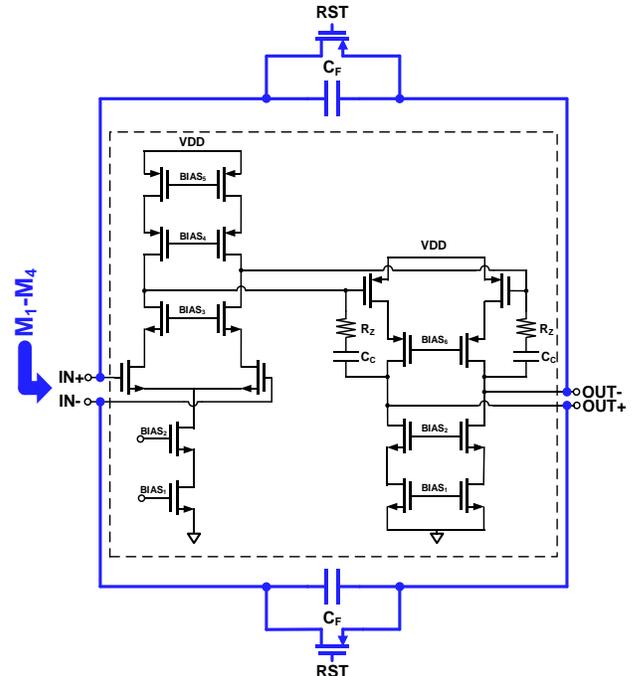

Fig. 13. Two-stage internally compensated OTA with the feedback network (Bias circuits are not shown).

The binary ($\pm1$) entries in the proposed THM based MACs permit: (i) half of the signal vector to be uniquely combined with the other half; (ii) easy realization through differential implementation without requiring any extra hardware; and (iii) easy scalability to a higher number of elements thus promising low-latency operation. A low-power source-follower buffer consuming 0.25 mA is used to drive each MAC output (OUT) for off-chip measurement.

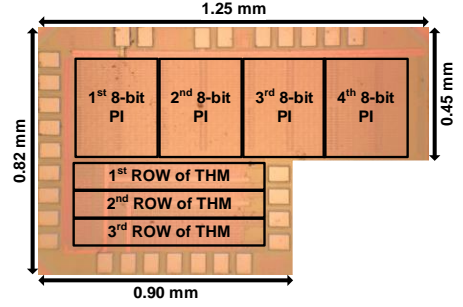

Fig. 14. Die micrograph in 65 nm CMOS.

## IV. MEASUREMENT RESULTS

The 4-element MIMO BB RX is implemented in a 65 nm CMOS process in an area of 0.65 and 0.9 mm$^2$ without and with pads, respectively, as shown in Fig. 14. The prototype is packaged in a Quad-Flat Package (QFP) enclosure to minimize parasitic bond wires. An SPI control port is used to set the on-chip digital phase-interpolation and time-interleaving. The delays between the RX antenna signals is implemented through two NI PXIe-5450 AWGs with four differential outputs and 145 MHz modulation BW as shown in the test setup in Fig. 15. Both single-tone and modulated signals are applied as inputs to characterize the SpICa.

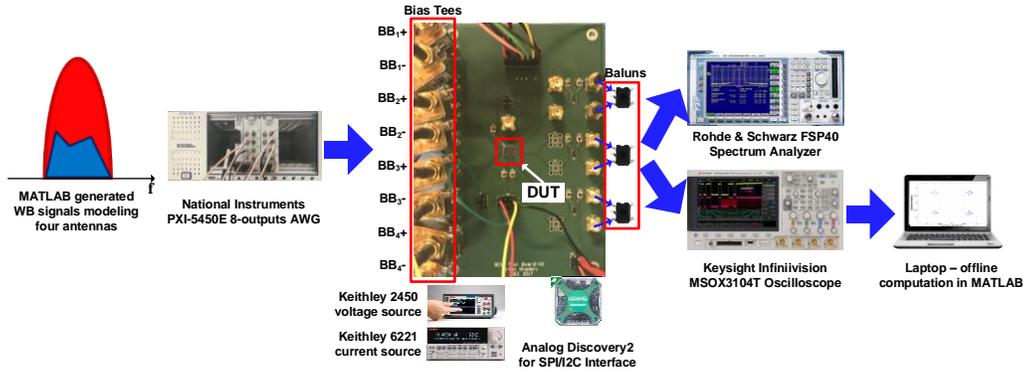

Fig. 15. Test Setup.

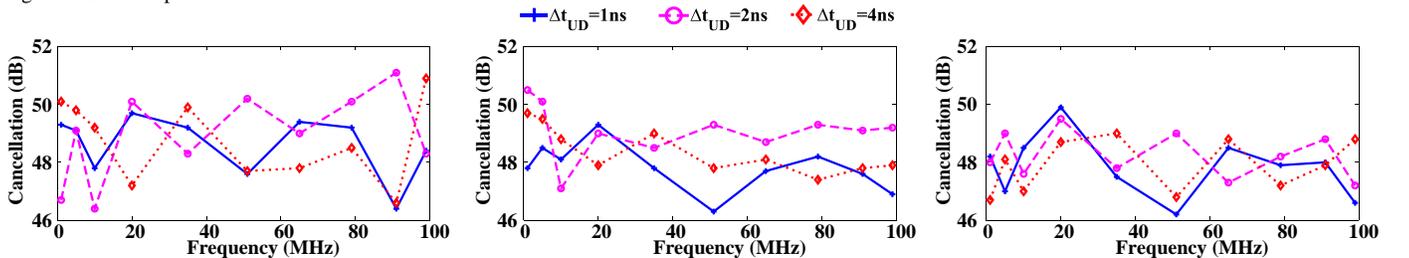

Fig. 16. Measurement results of the single-tone undesired signal SpICa performance vs. frequency for three cases of $\Delta t_{UD}$.

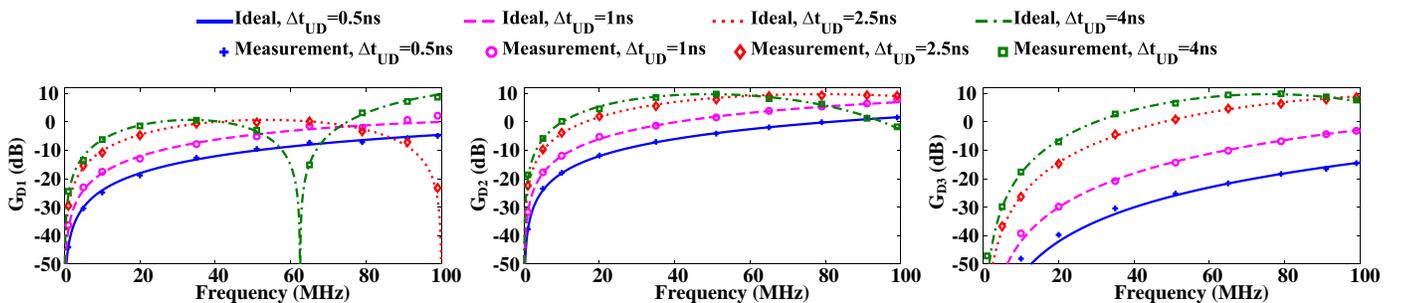

Fig. 17. Measurement and simulation results of the single-tone desired signal conversion gain vs. frequency for $\Delta t_D = 0$ and four cases of $\Delta t_{UD}$.

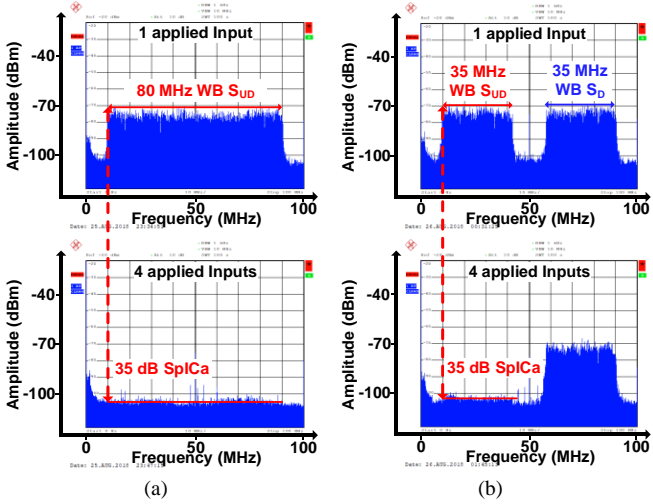

Fig. 18. SpICa measurement for a wide modulated BW undesired signal vs. frequency: (a) without any desired signal; and (b) with a wide modulated BW desired signal.

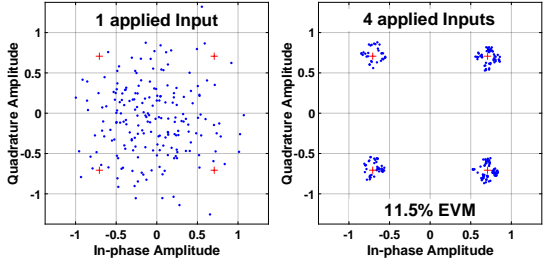

Fig. 19. Reconstructed constellation after measurement of a desired 4Mb/s QPSK signal in presence of a 12 dB stronger wide modulated BW interference.

Fig. 16 shows the measured SpICa performance for multiple AoA$_{UD}$ (corresponding to $\Delta t_{UD}$ = 1, 2, 4 ns) and a swept single-tone from 1 MHz (balun-limited) to 99 MHz. As it can be seen, the cancellation across the entire band is independent from AoA and >46 dB is achieved with 1.5× improvement in fractional BW over prior art [8].

Fig. 17 shows the desired signal conversion gains for the swept single-tone inputs from 1 MHz to 99 MHz at the outputs of each of the THM rows, matching closely with the ideal simulated conversion gains. The desired signal gain is defined as the ratio of output power when all inputs are applied, divided by the output power when only one input is applied, assuming all the desired inputs are in-phase and there are no delays between the inputs. This measurement is repeated for four different AoA$_{UD}$ (corresponding to $\Delta t_{UD}$ = 0.5, 1, 2.5, 4 ns). In all the cases, the gain response follows closely to the theoretical response as described in Section II-C.

Fig. 18(a) shows measured wide modulated BW SpICa performance for 80 MHz BW at the output of the THM's second row with one input (no cancellation) and four inputs (cancellation enabled), demonstrating >35 dB SpICa for a fractional-BW of 80% in the DC-100 MHz band (80 MHz is selected to visualize the plot easier and demonstrate the SpICa clearly). Similar performance is obtained when the cancellation is performed with other THM rows, confirming that the proposed architecture can be used for generating multiple versions of the desired signal simultaneously, while cancelling the undesired interference signals. This desired conversion gains can be equalized in the digital domain.

Fig. 18(b) shows the desired- and the undesired-signal spectrum (35 MHz within the DC-100 MHz band) with one and four inputs. In this measurement, AoA$_{UD}$=0° and AoA$_D$=45°. More than 35 dB SpICa across the entire BW of undesired signal is obtained again. Note that the BW of the desired- and the undesired-signal is reduced from 80MHz to 35MHz to visually show the cancellation on a 100 MHz plot. Although not demonstrated herein, our proposed approach can cancel two spatially distinct interferences.

Fig. 19 depicts SpICa performance for a wide modulated BW interferer 12 dB higher in power than the desired 4 Mb/s QPSK signal, with an offline-computed EVM of 11.5%. The SpICa data is sampled and stored using the digital sampling oscilloscope followed by digital equalization and EVM estimation in MATLAB. Limited memory in the sampling oscilloscope has restricted our measurements to QPSK modulation only.

The measured total power consumption of the analog implementation of the THM is 8 mW/100 MHz. Another 44 mW is consumed in clocking including the 0.8 GHz to 0.2 GHz quadrature phase generator, the four 8-bit PIs and the time-interleaving circuit. The clocking power can be decreased further to less than 5 mW for smaller resolutions (for example, the clocking power is <5 mW for 4-bit resolution). The clocking power can also be decreased if a reduced delay range is needed. The design constraints and power consumption associated with realization of a high-dynamic range ADC is thus significantly relaxed. With the demonstrated >35 dB wide modulated BW SpICa, the ADC dynamic range is significantly relaxed by nearly 6-b. Notably, a 1b reduction in ADC resolution leads to nearly 4× power savings [21] for ADCs limited by thermal noise. Additionally, the ADC power consumption can increase quadratically for high sampling frequencies as in [21].

Table 1 compares the proposed wide modulated BW SpICa to prior-art. The demonstrated SpICa of >35 dB over 80 MHz of wide modulated BW interferers is the highest ever reported to the best of our knowledge. The delay range-bandwidth product for the proposed TTD implementation is 1.5 (=15 ns×100 MHz).

Wide modulated BW interference separation is also demonstrated in [11], where a 500 kSps 64-QAM signal is separated from a 400 kSps 16-QAM signal, with an allocated bandwidth that can be calculated as 112.5 kHz and 135 kHz, respectively. In comparison, in our work, we demonstrate a cancellation of 80 MHz modulated bandwidth, an improvement of 592×. Measured P$_{1dB}$ compression point of the BB RX is 4.7 dBm with 10.6 dBm IIP3. Measured noise power at the THM output is 330 μVrms. We are currently investigating other phase interpolation techniques to significantly reduce the power consumption for future prototypes [22], [23].

TABLE I
PERFORMANCE SUMMARY AND COMPARISON TO PRIOR-ART SPATIAL CANCELLATION IMPLEMENTATIONS.

|  |  | JSSC2017 [8] | RFIC2016 [9] | JSSC2014 [10] | ISSCC2017 [11] | Proposed Work |
|---|---|---|---|---|---|---|
| # Channel | | 4 inputs, 4 outputs | 4 input, 1 output | 4 inputs, 1 output | 8 inputs, 8 outputs | **4 inputs, 3 outputs** |
| CMOS Tech. (nm) | | 65 | 65 | 65 | 65 | **65** |
| VDD (V) | | 1.2 | 1.3-1.5 | 1.2 | 1.2 | **1.0** |
| Resolution | | Phase: 6.5-b (3.8°) Amp: 3.9-b | 6-b I/Q | Phase: 3-b | 14-b | **8-b (5ps) Overall delay range: 15ns** |
| BB Power | | Not Available (RF+BB implementation) | Not Applicable (RF only implementation) | 36mW/40MHz[1] | 91µW/350kHz | **BB: 8mW/100MHz Clock: 44mW Total: 52mW** |
| Area (mm$^2$) | | 2.25 | 3.8 | 0.97 | 3.24 | **0.9** |
| P$_{IN1dB}$[2] (dBm) | | Not Available | Not Available | -5[3] | Not Available | **4.7[3]** |
| P$_{IIP3}$[2] (dBm) | | -29[3, 4, 5] | Not Available | 0-2.6 | Not Available | **10.6** |
| Noise Performance | | 3.4-5.8 dB[5] (Noise Figure) | 9.5 dB[5] (Noise Figure) | 3-6 dB (Noise Figure) | Not Available | **330 µV$_{rms}$ (Output-referred)** |
| SpICa Frequency | | 0.3-0.7GHz | 100MHz @10GHz | 40MHz @2.4GHz | 350kHz | **1-100MHz** |
| NB-SpICa[6] | Cancellation (dB) | 20 | 20 | < 38 | 84 | **46-51** |
|  | Range (MHz) | 320 | 100 | 40 | 0.35 | **99** |
| Modulated-SpICa | Cancellation (dB) | – | – | – | Not Available | **>35** |
|  | BW |  |  |  | 135 kHz | **80 MHz** |

[1] Does not include LO power, [2] NB-SpICa includes swept single-tone measurements, [3] Desired signal, [4] P$_{OIP3}$-Gain, [5] Cancellation enabled, [6] Including swept single-tone measurements.

## V. CONCLUSION

This work demonstrates a wide modulated BW spatial interference cancellation utilizing a discrete TD technique with Truncated Hadamard Matrix (THM) for a 4-element BB RX array. The proposed wide modulated BW SpICa implements a digital time-interleaver with 5 ps resolution and 15 ns range in a BW-, process- and supply-scalable implementation. The proposed architecture consumes less power (and area) compared to RF-based SpICa using a non-uniform sampling scheme with orthogonal THM implementation. Measurements demonstrate uniform cancellation across 80 MHz modulated interferer covering entire 100% modulated BW unlike prior-arts. The implemented spatial null BW is much larger than state-of-the art and represent the first demonstration of wide modulated BW signal null with 35 dB of rejection. The proposed wide modulated BW SpICa RX baseband augments the existing phased-array front-ends and can be easily adapted for even wider modulated BWs.

## ACKNOWLEDGEMENTS

The authors acknowledge the contributions of National Instruments for customized AWG. Access to CAD tools and technology was facilitated by MOSIS and CMC Microsystems. The authors are grateful to Profs. Deukhyoun Heo, Vishal Saxena and Akbar Sayeed for useful comments.